\documentclass[twoside,twocolumn,english,pre,showpacs]{revtex4-1}
\usepackage[T1]{fontenc}
\usepackage[latin9]{inputenc}
\usepackage{xcolor}
\usepackage{pdfcolmk}
\usepackage{amsmath}
\usepackage{amssymb}
\usepackage{graphicx}
\usepackage{esint}
\PassOptionsToPackage{normalem}{ulem}
\usepackage{ulem}

\makeatletter

\providecolor{lyxadded}{rgb}{0,0,1}
\providecolor{lyxdeleted}{rgb}{1,0,0}

 \@ifundefined{textcolor}{}
 {%
  \definecolor{BLACK}{gray}{0}
  \definecolor{WHITE}{gray}{1}
  \definecolor{RED}{rgb}{1,0,0}
  \definecolor{GREEN}{rgb}{0,1,0}
  \definecolor{BLUE}{rgb}{0,0,1}
  \definecolor{CYAN}{cmyk}{1,0,0,0}
  \definecolor{MAGENTA}{cmyk}{0,1,0,0}
  \definecolor{YELLOW}{cmyk}{0,0,1,0}
  }

\newcommand{\C}[1]{}
\newcommand{\bS}{\begin{subequations}}
\newcommand{\eS}{\end{subequations}}

\makeatother

\usepackage{babel}
\begin{document}
\global\long\def\Lp{L_{\parallel}}
\global\long\def\Ls{L_{\perp}}
\global\long\def\Lc{L_{\times}}
\global\long\def\xp{x_{\parallel}}
\global\long\def\xs{x}
\global\long\def\dx{\delta x}
\global\long\def\f{\psi}

\title{Strongly anisotropic non-equilibrium phase transition in Ising models
with friction}

\author{Sebastian Angst, Alfred Hucht, Dietrich E. Wolf}

\affiliation{Fakultät für Physik und CeNIDE, Universität Duisburg-Essen, D-47048
Duisburg}

\date{\today}
\begin{abstract}
The non-equilibrium phase transition in driven two-dimensional Ising
models with two different geometries is investigated using Monte Carlo
methods as well as analytical calculations. The models show dissipation
through fluctuation induced friction near the critical point. We first
consider high driving velocities and demonstrate that both systems
are in the same universality class and undergo a strongly anisotropic
non-equilibrium phase transition, with anisotropy exponent $\theta=3$.
Within a field theoretical \emph{ansatz} the simulation results are
confirmed. The crossover from Ising to mean field behavior in dependency
of system size and driving velocity is analyzed using crossover scaling.
It turns out that for all finite velocities the phase transition becomes
strongly anisotropic in the thermodynamic limit. 
\end{abstract}

\pacs{05.70.Ln, 68.35.Af, 05.50.+q, 05.70.Fh}

\maketitle

\section{Introduction}

The interest in magnetic contributions to friction due to spin correlations
has strongly increased in recent years. One interesting aspect is
the energy dissipation due to the formation of spin waves in a two-dimensional
Heisenberg model induced by a moving magnetic tip \cite{Fusco2008,Magiera2009,Magiera2009_2},
which can be of Stokes or Coulomb type depending on the intrinsic
relaxation time scales \cite{Magiera2011}. On the other hand, magnetic
friction occurs also in bulk systems moving relative to each other.
Kadau \textit{et al.} \cite{KadauHuchtWolf2008} used a two-dimensional
Ising model, cut into two halves parallel to one axis and moved along
this cut with the velocity $v$, to explore surface friction. The
motion drives the system out of equilibrium into a steady state, leading
to a permanent energy flux from the surface to the heat bath. This
model exhibits a non-equilibrium phase transition, which has been
investigated in several different geometries \cite{Hucht2009} by
means of analytical treatment as well as Monte Carlo (MC) simulations.
The critical temperature $T_{\mathrm{c}}$ of the considered models
depends on the velocity $v$ and has been calculated exactly for various
geometries in the limit $v\to\infty$. In this limit the class of
models show mean field-like critical behavior. Subsequent investigations
have been done in a variety of context, in particular for driven Potts
models \cite{Igloi2010} and for rotating Ising chains of finite length
\cite{Hilhorst2011}.

The nature of non-equilibrium phase transitions is still a field of
large interest, and simple models helping to explore this field are
seldom. A very famous example is the driven lattice gas (DLG) \cite{Katz1983,SchmittmannZia1995,Zia2010},
exhibiting a strongly anisotropic phase transition. Despite a lot
of similarities between the driven lattice gas and the Ising model
with friction, there is an important difference: The order parameter
is conserved in the former, while it is non-conserved in the latter
model. A further class of models characterized by non-equilibrium
phase transitions are sheared systems \cite{Chan1990,Onuki1997,Cirillo2005},
experimentally accessible within the framework of binary liquid mixtures. 

Like the driven lattice gas, the systems investigated in the following
exhibit a strongly anisotropic phase transition, which is investigated
by means of Monte Carlo simulations as well as a field theoretical
\textit{ansatz}.In addition, the case of finite velocities $v$ is
analyzed by means of crossover scaling, where a broad range of velocities
and system sizes are analysed. We show that for all $v>0$ the considered
models end up in the mean field class with strongly anisotropic correlations
as soon as the system size exceeds a velocity dependent crossover
length $\Lc(v)$. 

While a crossover behavior from Ising to mean field class occurs in
various thermodynamic systems such as ionic fluid \cite{Fisher1994,Gutkowski2001}
and spin systems with long-range interactions \cite{Luitjen1997},
to our knowledge such a crossover including a change from isotropic
to strongly anisotropic behavior has not been investigated in detail
until now. The paper is organized as follows: After introducing the
model and geometries, we determine the anisotropy exponent for $v=\infty$
using MC simulations as well as a field theoretical model. Then we
turn to finite velocities and present the crossover scaling analysis.
Finally we discuss our results.

\section{Models \label{sec:model}}

\begin{figure}
\begin{centering}
\includegraphics[scale=0.4]{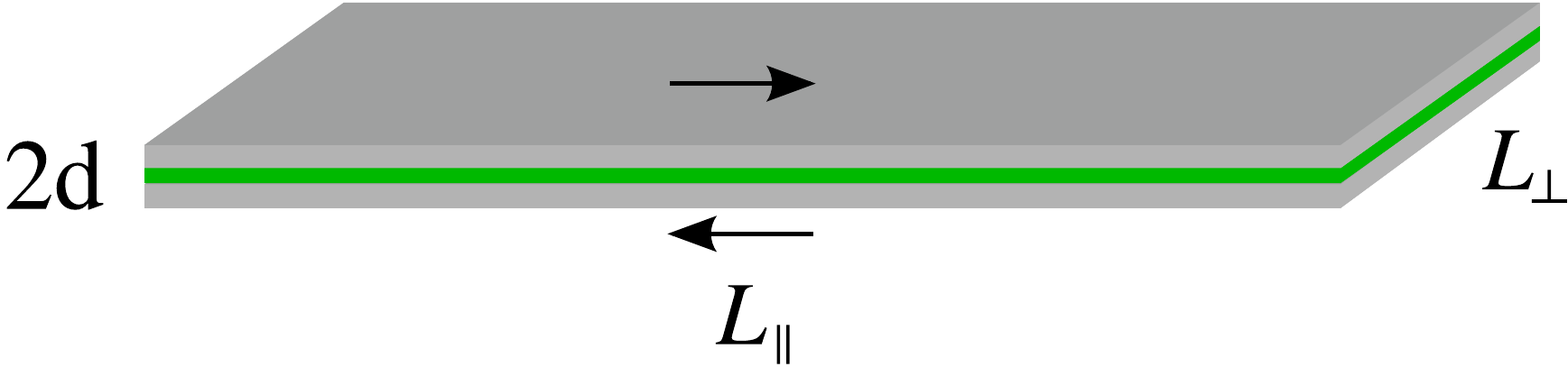}
\par\end{centering}

\begin{centering}
\medskip{}

\par\end{centering}

\begin{centering}
\includegraphics[scale=0.4]{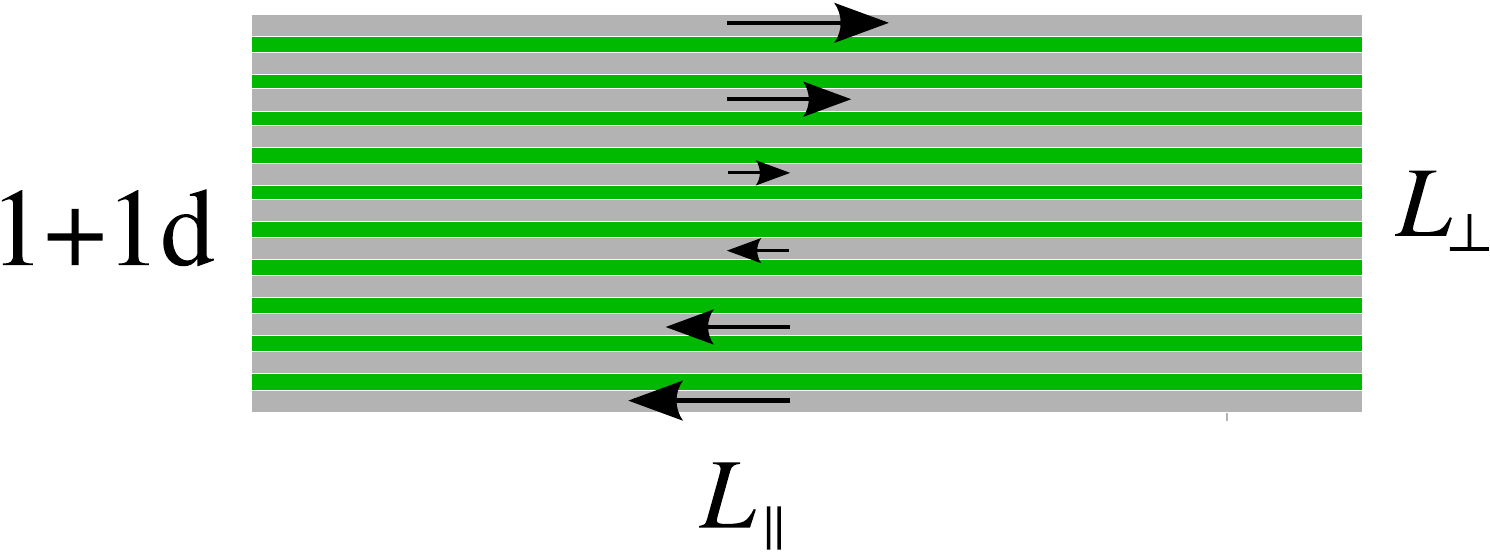}
\par\end{centering}

\caption{(Color online) The systems considered in this work. The gray regions
represent the magnetic systems, while the green (dark) regions are
the moving boundaries. The arrows indicate the motion of the subsystems.
\label{fig:model}}
\end{figure}

The systems considered in this work are denoted 2d and 1+1d and are
shown in Fig.\,\ref{fig:model} (for a classification see \cite{Hucht2009}).
The 2d system is a two-dimensional two-layer Ising model with $\Lp\times\Ls\times2$
lattice sites, where the two layers are moved relative to each other
along the parallel direction. Each lattice site carries one spin variable
$\sigma_{i,j,k}=\pm1$, and only nearest-neighbor interactions are
taken into account. Periodic boundary conditions are applied in both
planar directions, i.e., $\sigma_{i,j,k}=\sigma_{i+\Lp,j,k}=\sigma_{i,j+\Ls,k}$.
In order to simulate a finite velocity $v$ using Monte Carlo simulations
the upper sub-system is moved $v$ times by one lattice constant during
each random sequential Monte Carlo sweep (MCS). Since one MCS corresponds
to the typical time $t_{0}\approx10^{-8}\,\mathrm{s}$ a spin needs
to relax into the direction of its local Weiss field, and as the lattice
constant is of the order $a_{0}\approx10^{-10}\,\mathrm{m}$, the
velocity $v$ is given in natural units $a_{0}/t_{0}\approx1\,\mathrm{cm/s}$.

Instead of moving the two layers against each other, we reorder the
couplings between the subsystems with time to simplify the implementation
\cite{Hucht2009}. Introducing a time-dependent displacement 
\begin{equation}
\Delta(t)=vt,
\end{equation}
 which is increased by one after each $2L_{\parallel}L_{\perp}/v$
random sequential spin flip attempts, the Hamiltonian can be expressed
as 
\begin{equation}
\begin{split}\beta\mathcal{H}(t)=-K\sum_{i=1}^{L_{\parallel}}\sum_{j=1}^{L_{\vphantom{\parallel}\perp}}\sum_{k=0}^{1}\sigma_{i,j,k}(\sigma_{i+1,j,k}+\sigma_{i,j+1,k})\\
-K_{\mathrm{b}}\sum_{i=1}^{\Lp}\sum_{j=1}^{L_{\vphantom{\parallel}\perp}}\sigma_{i,j,0}\sigma_{i+\Delta(t),j,1},
\end{split}
\label{eq:Hamiltonian_2d}
\end{equation}
 with the reduced nearest neighbor coupling $K=\beta J$, the reduced
boundary coupling $K_{\mathrm{b}}=\beta J_{\mathrm{b}}$, and $\beta=1/k_{\mathrm{B}}T$.
In the following we assume $J=J_{\mathrm{b}}=1$.

 The critical temperature $T_{\mathrm{c}}(v)$ of the regarded systems
increases with $v$ and saturates for high velocities. In the limit
$v\rightarrow\infty$ an analytical calculation of the critical temperature
for the 2d geometry yield 
\begin{equation}
T_{\mathrm{c}}^{\mathrm{2d}}(\infty)=4.058782423...\label{eq:Tc_2d}
\end{equation}
for $J=J_{\mathrm{b}}=1$ \cite{Hucht2009}. The basic idea of the
analytic solution provides the approach for the implementation of
infinite velocity, which works as follows: the interaction partner
for a spin in the lower layer is chosen randomly from the same row
in the upper layer. Thus we can use Eq.~(\ref{eq:Hamiltonian_2d})
with a random value $1\le\Delta(t)\leq L_{\parallel}$.

The 1+1d system consists of a two-dimensional Ising model, where all
rows are moved relative to each other. The displacement $\Delta(t)=vt$
as well as the coupling $K_{\perp}$ is equal for all adjacent rows,
leading to the Hamiltonian 
\begin{equation}
\beta\mathcal{H}(t)=-\sum_{i=1}^{\Lp}\sum_{j=1}^{L_{\vphantom{\parallel}\perp}}K_{\parallel}\sigma_{i,j}\sigma_{i+1,j}+K_{\perp}\sigma_{i,j}\sigma_{i+\Delta(t),j+1}.\label{eq:Hamiltonian_1+1d}
\end{equation}
Again, periodic boundary conditions are applied in both directions,
where discontinuities in $\perp$ direction are avoided through the
homogeneous displacement $\Delta(t)$ \cite{Hucht2009}. The analytical
treatment at $v\to\infty$ gave the critical temperature 
\begin{equation}
T_{\mathrm{c}}^{1+1\mathrm{d}}(\infty)=1/\log\!\left({\textstyle {\frac{1}{2}\sqrt{3+\sqrt{17}}}}\right)=3.46591...\label{eq:Tc_1+1d}
\end{equation}
for $J_{\parallel}=J_{\perp}=1$ in this case \cite{Hucht2009}. Within
the scope of the 1+1d model the velocity $v$ corresponds to a shear
rate, which is ofter denoted as $\dot{\gamma}$ \cite{Gonnella2009,Winter2010}.
However, we will use the term velocity for both driving mechanisms
throughout this work.

\begin{figure}
\begin{centering}
\includegraphics[scale=0.25]{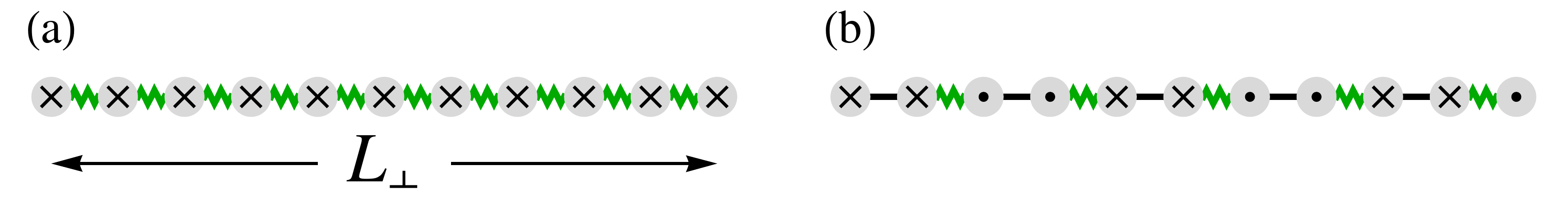}
\par\end{centering}

\centering{}\includegraphics[scale=0.25]{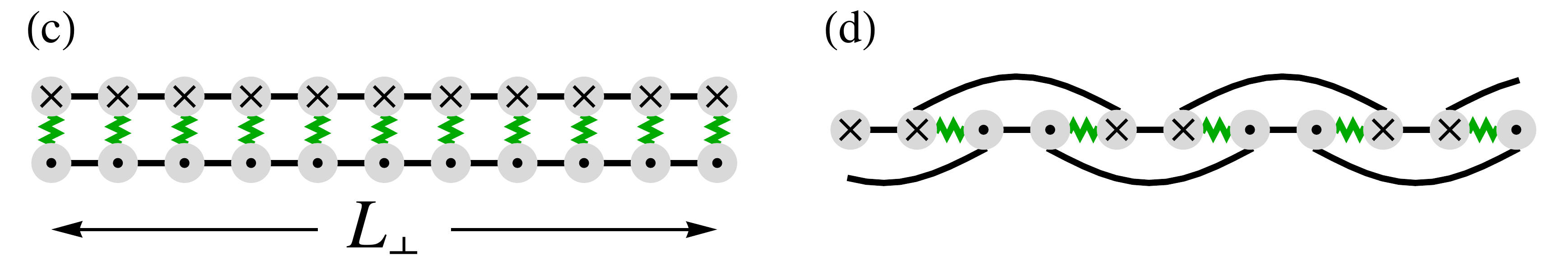} \caption{(Color online) Cross sections of the 1+1d (a) and the 2d model (c),
and slight modifications of both models ((b) and (d)). The grey circles
represent spin chains and the connecting lines substitute for the
coupling, where green wiggled lines stand for moving and black lines
for stationary couplings. Black crosses and dots indicate a motion
into and out of the plane, respectively. \label{fig:mapping}}
\end{figure}

In the following we argue that both systems show the same underlying
critical behavior. In order to emphasize the similarity, Fig.~\ref{fig:mapping}
illustrates slight variations of both models. First of all we start
with the 1+1d model (a) and change every second bond perpendicular
to the motion into a stationary bond. Additionally, we perform a transformation
that changes the homogeneous shear $\Delta(t)$ into an alternating
shift $\pm\Delta(t)$ of the double chains and reverses ($i\to-i$)
every second double chain, leading to the configuration in Fig.~\ref{fig:mapping}b.
These modifications do not change the critical behavior of the 1+1d
system, since still one-dimensional chains (now consisting of two
rows) are moved relative to each other. On the other hand, the cross
section of the 2d model can be visualized in a slightly different
way (see Fig.~\ref{fig:mapping}d) without altering the corresponding
Hamiltonian, Eq.~(\ref{eq:Hamiltonian_2d}). Since the next nearest
double chains in (b) are not moving relative to each other, the only
difference between (b) and (d) are the third nearest neighbor bonds
in (d), which are irrelevant at the critical point where long range
correlations dominate. Hence we conclude that both systems belong
to the same universality class. 

\begin{figure*}
\begin{centering}
\includegraphics[width=1\textwidth]{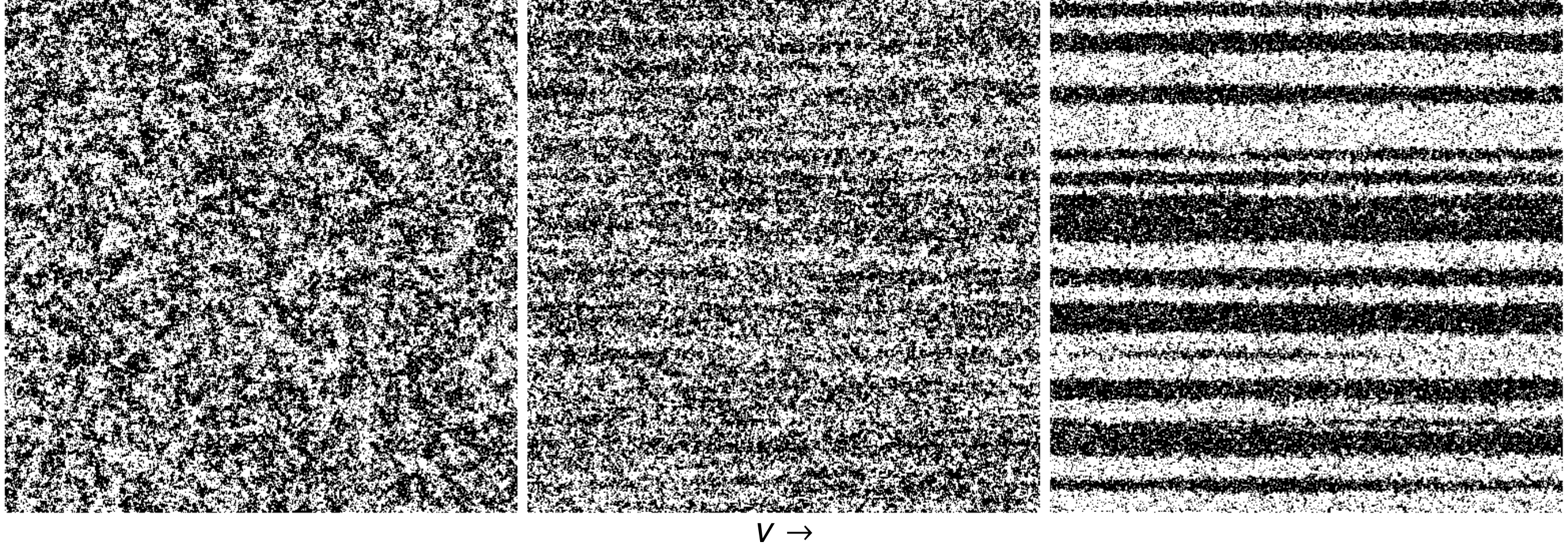}
\par\end{centering}

\centering{}\caption{Snapshots of one layer of the 2d model with $\Lp=\Ls=512$ and $J_{\parallel}=J_{\perp}=1$
at temperature $T=3.5$, which fulfills $T_{\mathrm{c}}^{\mathrm{2d}}(0)<T<T_{\mathrm{c}}^{\mathrm{2d}}(\infty)$.
We start with an equilibrium system at $v=0$ (left), set $v=\infty$,
and show the evolution at $t=42$ MCS (center) and $t=360$ MCS (right).
\label{fig:coarsening}}
\end{figure*}
Finally we mention that we must use the multiplicative rate 
\begin{equation}
p_{\text{flip}}(\Delta E)=e^{-\frac{\beta}{2}(\Delta E-\Delta E_{\text{min}})},
\end{equation}
with $\Delta E_{\text{min}}=\min(\{\Delta E\})$ to reproduce the
critical temperatures, Eqs.(\ref{eq:Tc_2d}, \ref{eq:Tc_1+1d}), in
simulations, for a discussion see \cite{Hucht2009}.

\section{Results \label{sec:results}}

In order to illustrate symptomatic features of both systems, Fig.~\ref{fig:coarsening}
shows a sequence of spin configurations of one layer of the $2$d
system (note that the same characteristics are observed in the 1+1d
system). On the left hand side an equilibrated system at $T=3.5$
well above the critical temperature of the non-moving system, $T_{\mathrm{c}}^{\mathrm{2d}}(0)=3.20755(5)$
\cite{Lipowski1993}, is presented. Shortly after starting the motion
stripe-like domains arise, spanning the whole system parallel to the
motion. The stripes are rather stable, but are nonetheless transient,
since they grow in time until the system ends up in a homogeneously
magnetized state. The evolution in Fig.~\ref{fig:coarsening} is
an example for a velocity driven phase transition already described
in \cite{KadauHuchtWolf2008,Hucht2009}, which is triggered by the
onset of the motion and the associated increase of the critical temperature.
The circumstances are comparable to a quench, which is characterized
by a temperature decrease below $T_{\mathrm{c}}$. After a quench
a coarsening of domains is observed, whereas the growth of the domains
can be described by a power law (e.g.~\cite{Bray1994,Puri2004}).
Domain growth in systems exhibiting a strongly anisotropic phase transition,
e.g., the DLG model, is also a well investigated subject \cite{Yeung1992,SchmittmannZia1995,Hurtado2002}.
The corresponding time evolution of spin configurations are similar
to those shown in Fig.~\ref{fig:coarsening}, leading to the assumption
that the 2d and the 1+1d geometries are also characterized by strongly
anisotropic correlations, which is shown in the following section.

\subsection{Determination of $\theta$ in the limit $v\rightarrow\infty$ \label{subsec:anis}}

A strongly anisotropic phase transition is characterized by a correlation
length $\xi_{\mu}$ which diverges with direction dependent critical
exponents $\nu_{\mu}$ at the critical point %
\footnote{Throughout this work, the symbol $\sim$ means ``asymptotically equal''
in the respective limit, e.\,g., $f(L)\sim g(L)\Leftrightarrow\lim_{L\rightarrow\infty}f(L)/g(L)=1.$
Note that the variable $t$ is used for the reduced temperature throughout
the rest of this work.%
},
\begin{equation}
\xi_{\mu}(t)\stackrel{{\scriptscriptstyle t>0}}{\sim}\hat{\xi}_{\mu}t^{-\nu_{\mu}},\label{eq:xi}
\end{equation}
with direction $\mu=\{\perp,\parallel\}$ and reduced critical temperature
$t=T/T_{\mathrm{c}}-1$. Defining the anisotropy exponent \cite{Binder1990,Henkel1999,Hucht2002}
\begin{equation}
\theta=\frac{\nu_{\parallel}}{\nu_{\perp}},\label{eq:theta}
\end{equation}
we find 
\begin{equation}
\xi_{\parallel}(t)/\xi_{\perp}^{\theta}(t)\sim\hat{\xi}_{\parallel}/\hat{\xi}_{\perp}^{\theta}\label{eq:xi_durch_xi}
\end{equation}
independent of $t$. Isotropic scaling takes place for $\theta=1$
and strongly anisotropic scaling is implied by $\theta\ne1$. Several
models with strongly anisotropic behavior where studied in the past.
Examples are Lifshitz points as present in the anisotropic next nearest
neighbor Ising (ANNNI) model \cite{Selke1988,HenkelPleimling2001},
the non-equilibrium phase transition in the DLG \cite{SchmittmannZia1995},
the two-dimensional dipolar in-plane Ising-model \cite{Hucht2002}.
Furthermore, strongly anisotropic behavior usually occurs in dynamical
systems, where the parallel direction can be identified with time
and the perpendicular direction(s) with space \cite{Henkel1999,HayeHabil2000}.
In the latter case the anisotropy exponent $\theta$ corresponds to
the dynamical exponent $z$.

The knowledge of the anisotropy exponent is essential and necessary
for appropriate simulations of strongly anisotropic systems. To avoid
complicated shape effects it is required to keep the generalized aspect
ratio \cite{Binder1990,Henkel1999,Hucht2002}

\begin{equation}
\rho=\frac{\Lp/\hat{\xi}_{\parallel}}{(\Ls/\hat{\xi}_{\perp})^{\theta}}\label{eq:rho}
\end{equation}
fixed, which requires the knowledge of $\theta$. We will show in
the following that the limit $\rho\to0$ simplifies the analysis for
infinite velocity $v$ and turns out to be essential at finite $v$.

\begin{figure*}
\hfill{}\includegraphics[scale=0.37]{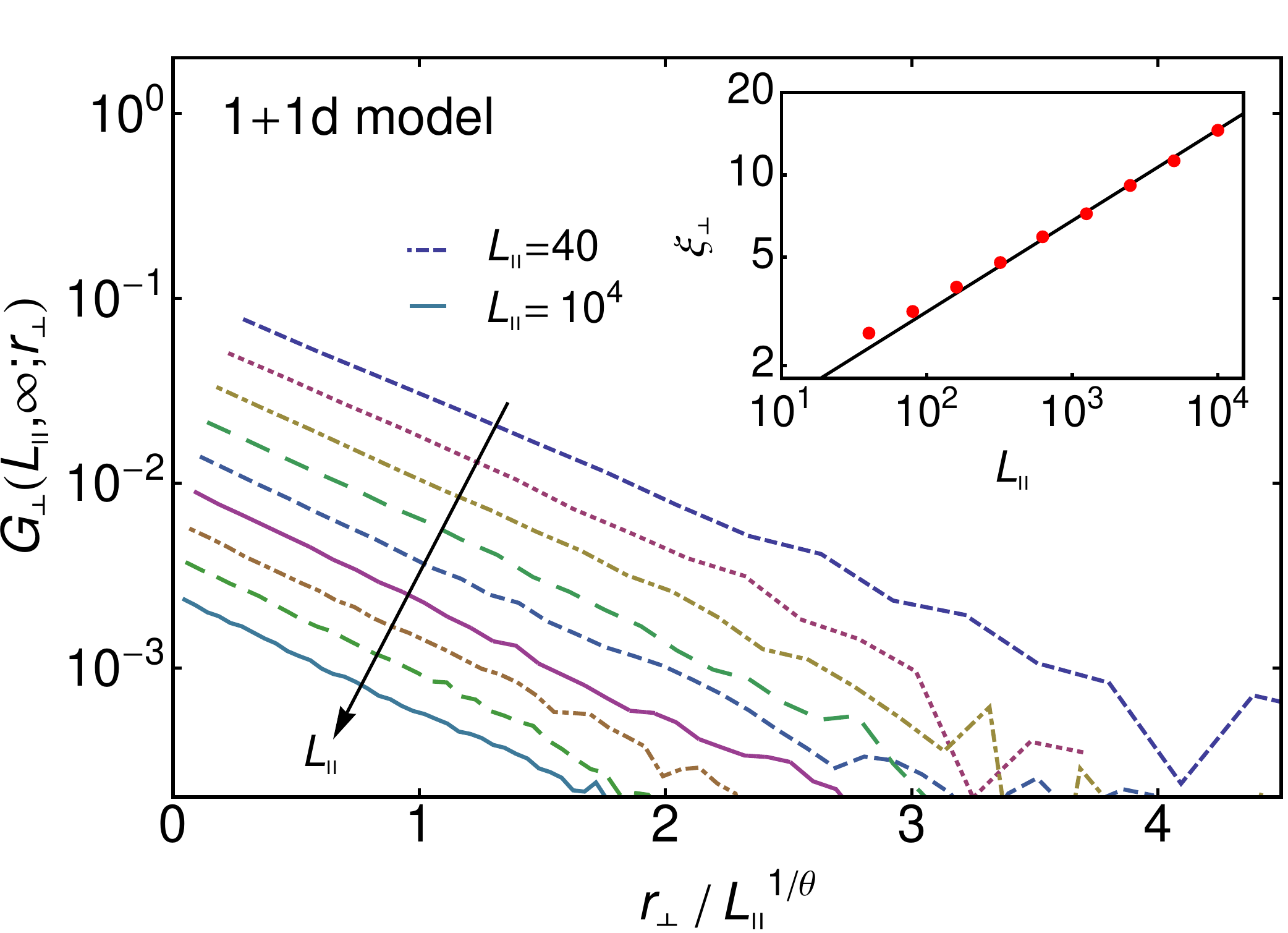}\hfill{}\includegraphics[scale=0.37]{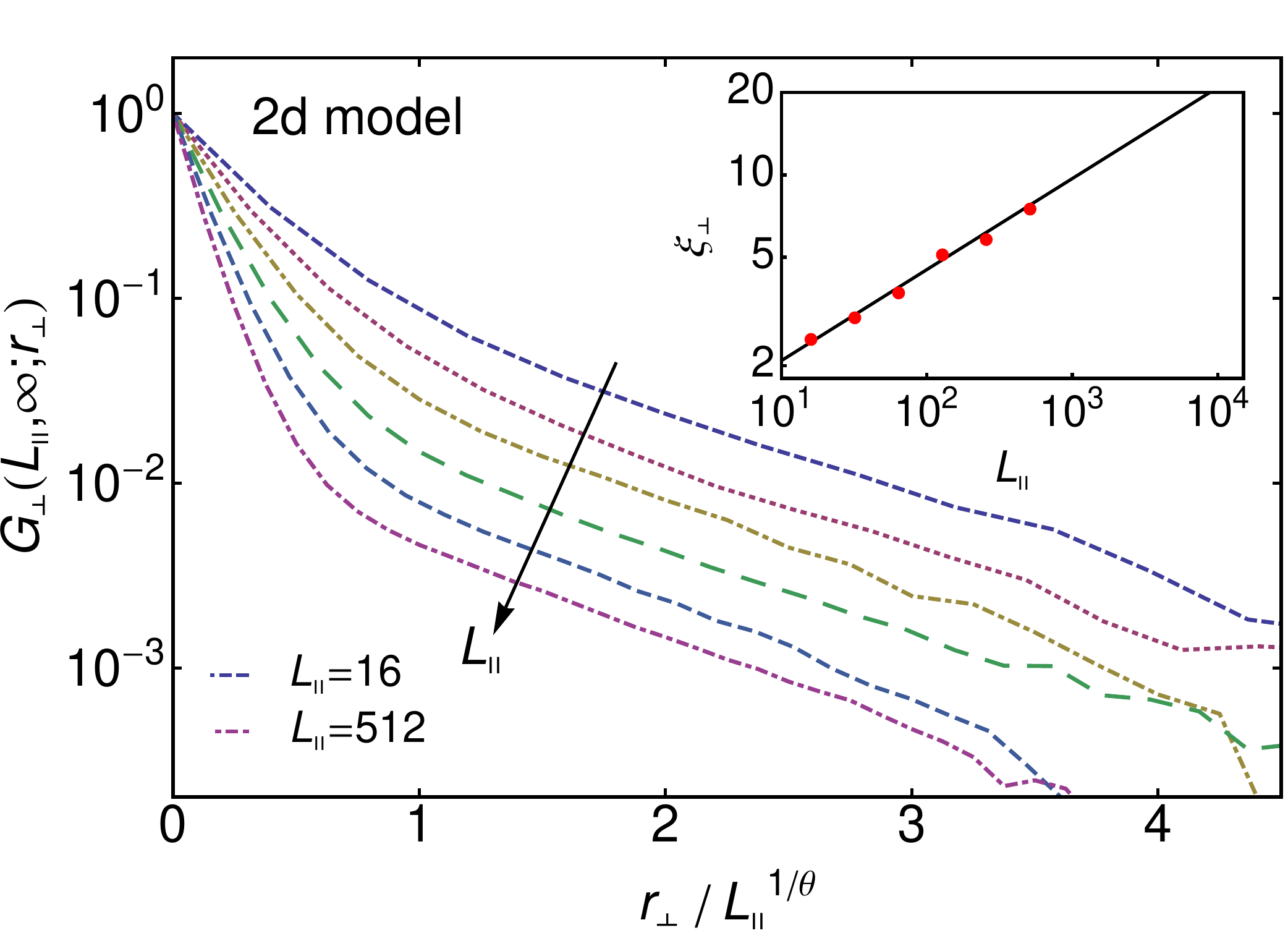}\hfill{}

\caption{(Color online) Rescaled correlation function $G_{\perp}(\Lp,\infty;r_{\perp})$
at criticality for both models for varying system extensions, $\Lp=\{40,80,160,320,625,1250,2500,5000,10000\}$
(1+1d) and $\Lp=\{16,32,64,128,256,512\}$ (2d), respectively. The
insets show $\xi_{\perp}(\Lp)$ whereby we yield $\xi_{\perp}$ by
fitting an exponential function to the long-range part of $G_{\perp}(\Lp,\infty;r_{\perp})$.
The solid line is a power law with exponent $\theta^{-1}=1/3$ as
predicted by the field theoretical analysis (see text). \label{fig:corr}}
\end{figure*}
We first discuss the case $v\to\infty$ and always assume criticality,
$t=0$. In order to determine the anisotropy exponent $\theta$ we
calculate the perpendicular correlation function $G_{\perp}(\Lp,\Ls;r_{\perp})=\langle\sigma_{i,j}\sigma_{i,j+r_{\perp}}\rangle$
between spins at distance $r_{\perp}$ in cylinder geometry $\Ls\rightarrow\infty$
(leading to $\rho\to0$), and thereby gain the correlation length
$\xi_{\perp}(\Lp)$ through 
\begin{equation}
G_{\perp}(\Lp,\infty;r_{\perp})\sim\hat{G}_{\perp}(\Lp)\, e^{-r_{\perp}/\xi_{\perp}(\Lp)},\label{eq:Gperp}
\end{equation}
where the prefactor $\hat{G}_{\perp}(\Lp)$ is shown to be proportional
to $\Lp^{-2/3}$ in Appendix~\ref{sec:appendix}. Approaching the
critical point within the given geometry, the correlation length $\xi_{\parallel}(t)$
is limited by $\Lp$, and using Eq.~(\ref{eq:xi_durch_xi}) this
leads to the relation 
\begin{equation}
\xi_{\perp}(\Lp)\sim A_{\perp}\Lp^{1/\theta}\label{eq:Lp_vs_xs}
\end{equation}
with non-universal amplitude $A_{\perp}$ \cite{HenkelSchollwock2001,Hucht2002}.
Measuring the correlation length $\xi_{\perp}$ in dependency of the
parallel extension $\Lp$ allows us to determine the anisotropy exponent
$\theta$.

In the simulations, the limit $\Ls\rightarrow\infty$ is implemented
by the condition $\Ls/\xi_{\perp}\gtrsim10$. This is sufficient to
keep the sytematic errors in $G_{\perp}$ smaller than the statistical
error $\epsilon=10^{-3}$ adequate to calculate $\xi_{\perp}$. From
$\epsilon$ we can determine the required system sizes via $\Ls/\xi_{\perp}=-2\log[\epsilon/\hat{G}_{\perp}(\Lp)]$,
where the factor 2 accounts for the periodic boundary conditions.
As $\hat{G}_{\perp}\approx0.1$ for $\Lp=40$ and $\hat{G}_{\perp}\approx0.02$
for $\Lp=10^{4}$ for the 1+1d model (see Fig.~\ref{fig:corr}(left))
we yield $\Ls/\xi_{\perp}\approx10$ for $\Lp=40$ and $\Ls/\xi_{\perp}\approx0.7$
for $\Lp=10^{4},$ meaning that for large systems a much smaller value
of $\Ls/\xi_{\perp}$ would be sufficient. 

Fig.~\ref{fig:corr} displays the correlation functions for both
models. For the 1+1d case these correlations are purely exponential
also at short distances, since the coupling in $\perp$ direction
is mediated through fluctuating fields \cite{Hucht2009}, leading
to dimensional reduction to an effectively one-dimensional system.
The resulting correlation length $\xi_{\perp}$ is shown in the inset
of Fig.~\ref{fig:corr}(left). The growth of $\xi_{\perp}(\Lp)$
follows a power law with exponent $\theta^{-1}=1/3$ and with prefactor
\begin{equation}
A_{\perp}^{\mathrm{1+1d}}=\lim_{\Lp\to\infty}\Lp^{-1/3}\xi_{\perp}^{\mathrm{1+1d}}(\Lp)=0.68(2),
\end{equation}
indicated as a black line. 

In the case of the 2d model (right figure in Fig.\ref{fig:corr})
we find two regions with different characteristics. The short-distance
correlations are affected by the $\perp$ nearest-neighbor interactions
within the planes which are not present in the 1+1d model. These correlations
decay with a correlation length of the order $\xi_{\perp}^{\mathrm{eq}}[T_{\mathrm{c}}^{\mathrm{2d}}(\infty)]\approx1$.
For large distances the correlations crossover to an exponential behavior.
The exponential correlations are propagated by the fluctuations of
stripe-like domains. The analysis yields 
\begin{equation}
A_{\perp}^{\mathrm{2d}}=\lim_{\Lp\to\infty}\Lp^{-1/3}\xi_{\perp}^{\mathrm{2d}}(\Lp)=0.94(3)
\end{equation}
in this case. 

From the anisotropy exponent $\theta=3$ we can derive the correlation
length exponents $\nu_{\parallel}=3/2$ and $\nu_{\perp}=1/2$ using
the generalized hyper-scaling relation 
\begin{equation}
2-\alpha=2\beta+\gamma=\nu_{\parallel}+(d-1)\nu_{\perp},
\end{equation}
with $d=2$ and mean field exponents $\alpha=0$, $\beta=1/2$, and
$\gamma=1$, whose validity has been demonstated in \cite{Hucht2009}
by a mapping onto a mean field equilibrium model.

The calculatation of $\theta$ in the limit $v\to\infty$ is done
within a one-dimensional Ginzburg-Landau-Wilson (GLW) field theory
\cite{BrezinZinnJustin1985}. For $v\to\infty$ it was shown in Ref.~\cite{Hucht2009}
that the 1+1d model can be mapped onto an \emph{equilibrium} system
consisting of one-dimensional chains that only couple via fluctuating
magnetic fields. Due to the stripe geometry with short length $\Lp$
and the periodic boundary conditions in $\parallel$ direction the
magnetization is homogeneous in $\parallel$ direction, and parallel
correlations are irrelevant. Hence we can use the zero mode approximation
in this direction. However, it is necessary to include a term representing
the interaction between adjacent spin chains. This can be expressed
by the square of the spatial derivative of the magnetization in the
direction $\perp$ to the motion. Hence the minimal GLW model to describe
this strongly anisotropic mean field system is given by 
\begin{equation}
\beta\mathcal{H}=\Lp\int_{0}^{\Ls}\!\!\!\!\mathrm{d}\xs\left(\frac{t}{2}m(\xs)^{2}+\frac{1}{2}m'(\xs)^{2}+\frac{u}{4!}m(\xs)^{4}\right)\label{eq:Hamiltonian_GLW}
\end{equation}
with phenomenological parameters $t$ and $u$, where $m(\xs)$ represents
the magnetization of the spin chain at $\perp$ coordinate $\xs$.
Eq.~(\ref{eq:Hamiltonian_GLW}) corresponds to the Hamiltonian used
for the description of a cylinder-like spin system, which is infinite
along one dimension, and finite and periodic in $d-1$ dimensions
\cite{BrezinZinnJustin1985}. The partition function of Eq.~(\ref{eq:Hamiltonian_GLW})
can be mapped onto a one-dimensional Schrödinger equation in a quartic
anharmonic oscillator potential using a rescalation, which yields
the critical exponents $\nu_{\parallel}=3/2$ and $\theta=3$. The
detailed derivation is given in Appendix~\ref{sec:appendix}.

\subsection{Crossover scaling at finite velocities}

\begin{figure*}
\centering{}\hfill{}\includegraphics[scale=0.59]{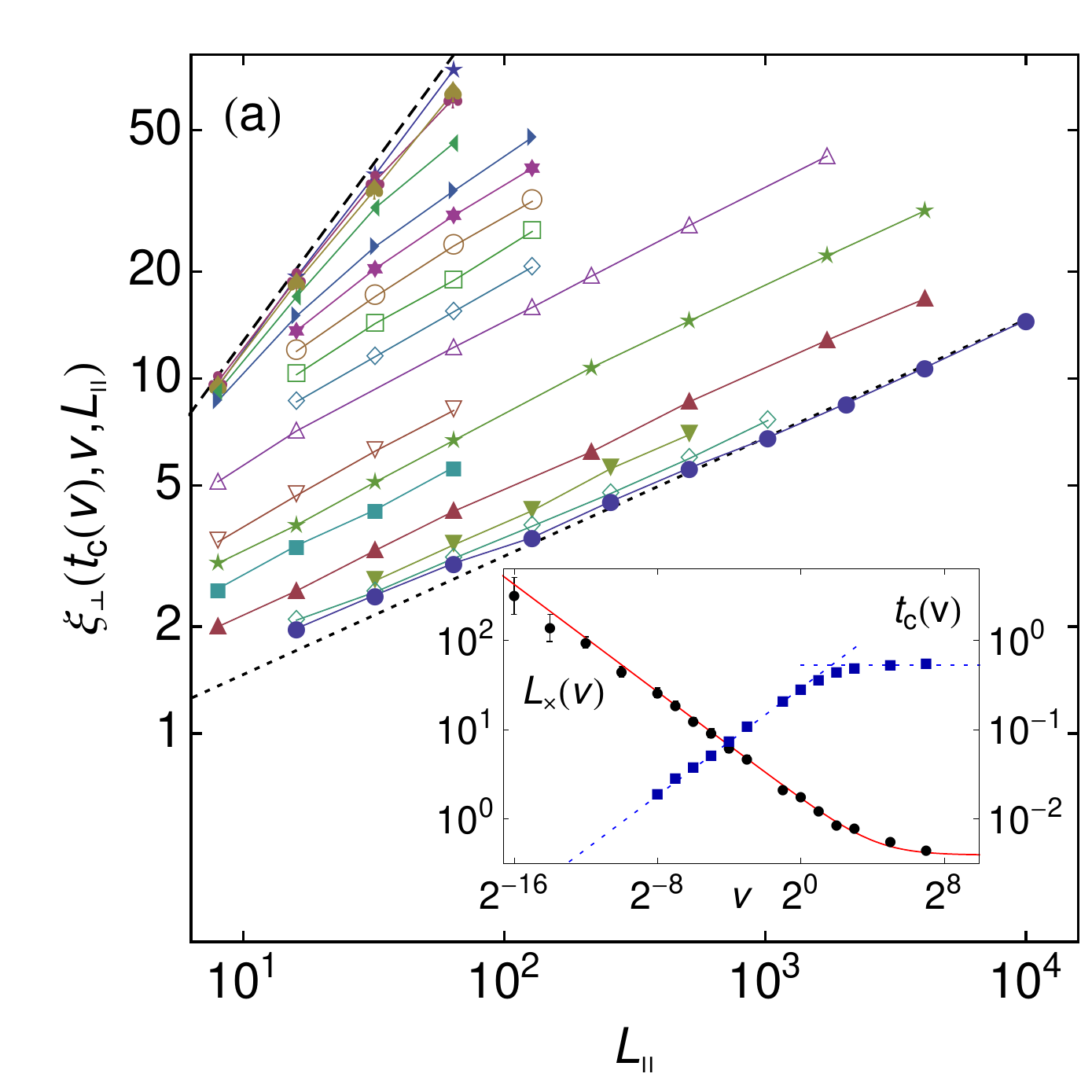}\hfill{}\includegraphics[scale=0.6]{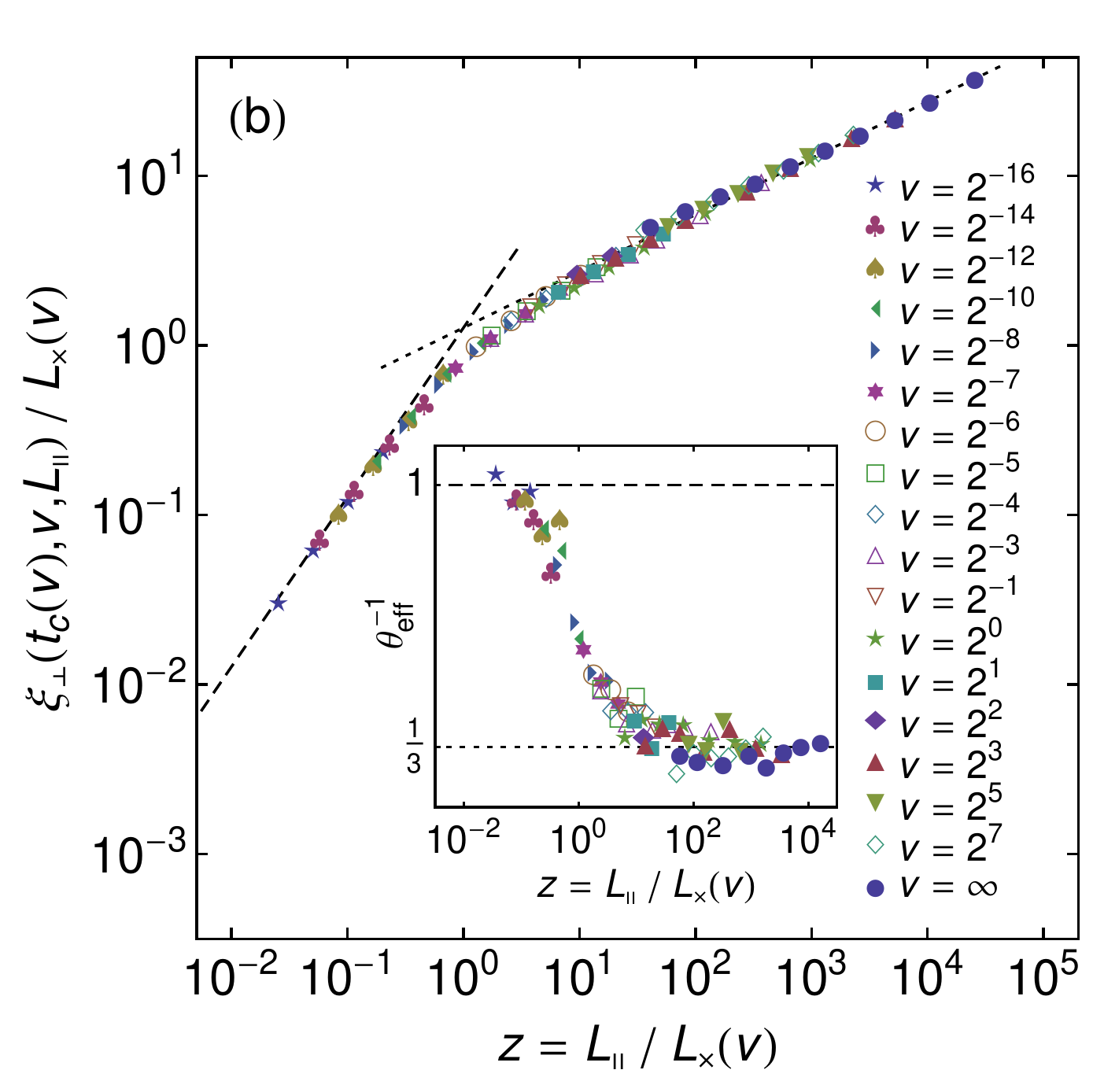}\hfill{}
\caption{(Color online) Velocity dependent crossover behavior in the 1+1d case.
Both pictures show log-log plots of the correlation length $\xi_{\perp}(t_{\mathrm{c}}(v),v,\Lp)$
as function of the system size $\Lp$ at reduced critical temperature
$t_{c}(v)$ for a broad range of different velocities $v$. The dashed
line is the analytically known Ising limit $\xi_{\perp}(0,0,\Lp)/\Lp\sim4/\pi$
valid for $v\to0$ \cite{Cardy1984}, while the dotted line has slope
$\theta^{-1}=1/3$. The left figure shows the unscaled data and the
inset displays the rescaling factor $\Lc(v)$ for different velocities
$v$ (black dots, see text) and a function approximating the data
given in Eq.~(\ref{eq:Lcross(v)}) (red solid line), as well as the
reduced critical temperature $t_{c}(v)$ (blue squares) together with
its asymptotes, Eqs.~(\ref{eq:Tc_1+1d},\ref{eq:tc(v)}). The right
figure displays the same data rescaled with the crossover length $\Lc(v)$,
leading to an excellent data collapse. The inset shows the crossover
of the effective anisotropy exponent $\theta_{\mathrm{eff}}$ from
$\theta_{\mathrm{eff}}=1$ (Ising, dashed line) to $\theta_{\mathrm{eff}}=3$
(MF, dotted line). \label{fig:crossplot}}
\end{figure*}
We now turn to finite velocities. The following analysis is exemplarily
done for the 1+1d model, but as stated above, both models belong to
the same universality class and similar results are expected for the
2d model. As we expect a crossover from an isotropic Ising model with
$\theta=1$ to a strongly anisotropic system with $\theta=3$, we
must be careful with the system geometry: We cannot use a fixed finite
generalized aspect ratio $\rho$, Eq.~(\ref{eq:rho}), in the simulations,
as $\theta$ is not constant. The only possible choice is $\rho\to0$
(or $\rho\to\infty$), where the $\theta$-dependency drops out. 

We consider the correlation length $\xi_{\perp}(t_{\mathrm{c}}(v),v,\Lp)$
at reduced critical temperature 
\begin{equation}
t_{\mathrm{c}}(v)=\frac{T_{\mathrm{c}}(v)}{T_{\mathrm{c}}(0)}-1,\label{eq:tc(v)_def}
\end{equation}
where $T_{\mathrm{c}}(0)=2/\log(\sqrt{2}+1)$. $t_{\mathrm{c}}(v)$
is calculated via a finite-size scaling analysis of the perpendicular
correlation length (not shown). As this procedure becomes inaccurate
for small velocities $v<2^{-8}$, we calculate the critical temperature
according to 
\begin{equation}
t_{\mathrm{c}}(v)\stackrel{{\scriptstyle v\to0}}{\sim}\hat{c}\, v^{\phi}\label{eq:tc(v)}
\end{equation}
with $\hat{c}=0.29(1)$ in these cases, where we assume $\phi=1/2$
in agreement with the literature \cite{SchmittmannZia1995,Gonnella2009,Winter2010}.
The results are shown in the inset of Fig.~\ref{fig:crossplot}a.

Fig.~\ref{fig:crossplot}a shows the unscaled data, which gives evidence
that the correlation length of systems moved at high velocities $v$
are well described by the exponent $\theta=3$ (dotted line), whereas
for low velocities $v\lesssim2^{-12}$ effectively the Ising exponent
$\theta=1$ (dashed line) holds for the simulated system sizes $\Lp$.
The curvature of the data of intermediate velocities suggest the crossover.
As a data collapse on the analytical known \cite{Cardy1984} relation
$\xi_{\perp}(0,0,\Lp)/\Lp\sim A^{\mathrm{eq}}=4/\pi$ (dashed line
in Fig.~\ref{fig:crossplot}) has to be obtained in the limit $v\rightarrow0$,
both axes must be rescaled by the same factor $\Lc(v)$. This \emph{crossover
length} can be determined by applying the following method: We start
with plotting the correlation length in the mean field limit $\xi_{\perp}(t_{\mathrm{c}}(\infty),\infty,\Lp)$.
Then we subsequently add the data for smaller $v$ by rescaling $\xi_{\perp}$
and $\Lp$ with $\Lc^{-1}(v)$, which shifts the points parallel to
the dashed line, until a data collapse is obtained (see Fig.~\ref{fig:crossplot}b).
This procedure works quite accurate for velocities $v\gtrsim2^{-3}$,
only at very small $v\lesssim2^{-12}$ the errors in $\Lc(v)$ grow
due to the fact that we just shift the data along the dashed line.
The resulting crossover length $\Lc(v)$ is pictured as black dots
in the inset of Fig.~\ref{fig:crossplot}a. The behavior of $\Lc(v)$
is analogous to the velocity dependency of other quantities like the
critical temperature or the energy dissipation, which are characterized
by a power law for $v\ll1$ and a saturation for $v\gg1$. 

\begin{figure*}
\centering{}\includegraphics[width=1\textwidth]{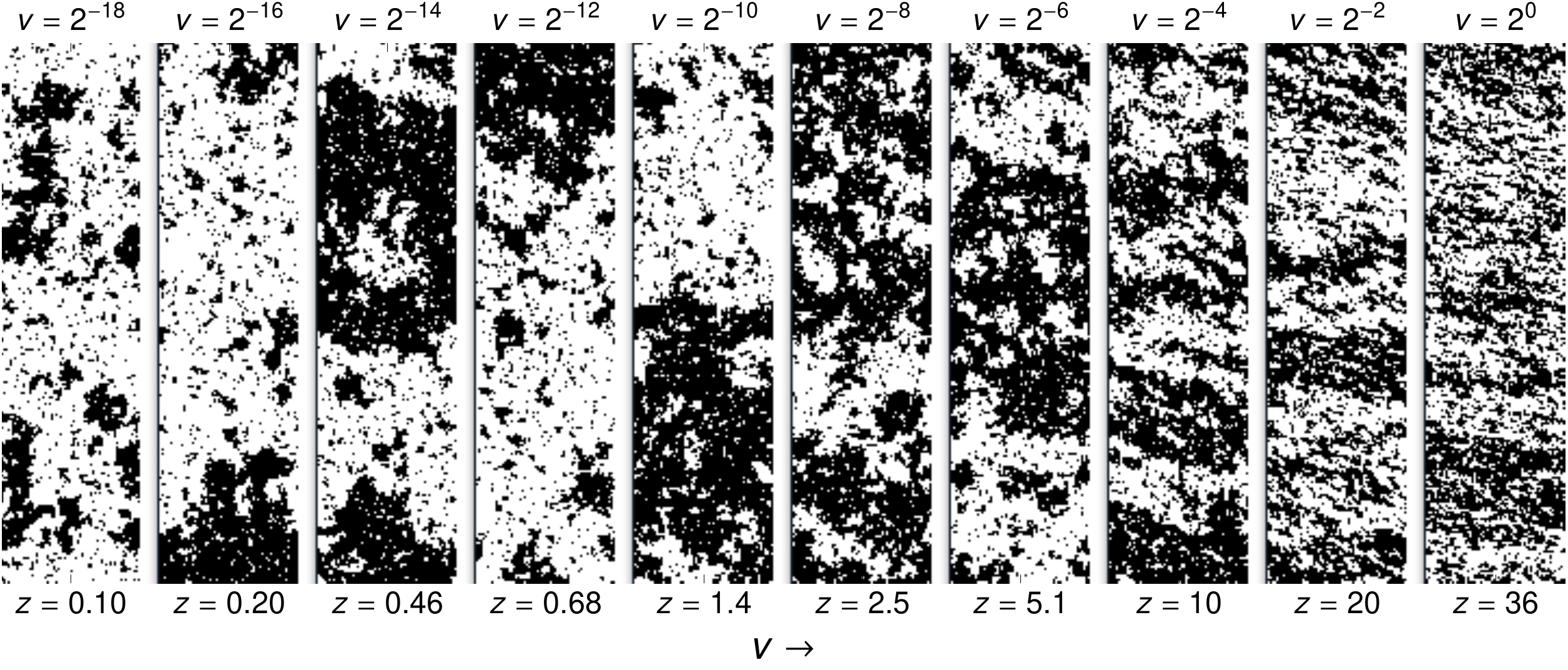} \caption{Typical spin configurations of the critical 1+1d system for $\Lp=64$
and different velocities $v=2^{-18},\ldots,1$. $z=\Lp/\Lc(v)$ denotes
the crossover scaling variable (see text). The critical domains are
isotropic and Ising-like for $z\ll1$ and become anisotropic for $z\gtrsim1$.
\label{fig:spinpic}}
\end{figure*}

We conclude that for all finite velocities $v>0$ the critical behavior
changes from Ising type to mean field type at a velocity dependent
crossover length $\Lc(v)$ approximately given by 
\begin{equation}
\Lc(v)\approx\left(\frac{A_{\perp}^{\mathrm{1+1d}}}{A^{\mathrm{eq}}}\right)^{3/2}\sqrt{1+\frac{v_{\times}}{v}}\label{eq:Lcross(v)}
\end{equation}
(solid red curve in the inset of Fig.~\ref{fig:crossplot}a), where
the velocity is measured in units $10^{-8}\,\mathrm{m/s}$ and the
size in $10^{-10}\,\mathrm{m}$. The velocity independent prefactor
was added to shift the crossover point, i.e., the intersection of
the asymptotes, to $z=1$. The saturation of $\Lc$ at $v_{\times}=18(2)$
results from the lattice cut-off, as $\Lc(v_{\times})\approx1$. The
inset in Fig.~\ref{fig:crossplot}b shows the effective exponent
$\theta_{\mathrm{eff}}$, obtained from the logarithmic derivative
\begin{equation}
\theta_{\mathrm{eff}}^{-1}=\frac{\partial\log\xi_{\perp}}{\partial\log\Lp},
\end{equation}
whose value changes from $\theta_{\mathrm{eff}}=1$ (Ising, isotropic)
to $\theta_{\mathrm{eff}}=3$ (MF, strongly anisotropic). Note that
we verified the mean field exponents for $v\gtrsim1/8$ with finite-size
scaling methods and also found good agreement of the scaling function
with the universal finite-size scaling function \cite{GrunbergHucht2004}(not
shown). In order to illustrate the change of the critical behavior,
Fig.~\ref{fig:spinpic} shows typical critical spin configurations
for different values of the crossover scaling variable $z=\Lp/\Lc(v)$.

We are now able to compare our results with the literature. If the
crossover scaling variable $z\ll1$ Ising-like behavior occurs, whereas
for $z\gg1$ mean field exponents and strongly anisotropic correlations
are expected. In experiments \cite{Onuki1997}, even slow shear rates
of the order of $10^{-4}$ (in natural units $t_{0}^{-1}$, where
now $t_{0}$ is the time scale of the fluid dynamics), lead to a crossover
length $L_{\times}\lesssim100$ and, as the typical system size is
large wrt. the atomic distances, give $z\gg1$, indicating that experimental
data are always obtained in the mean field limit. 

In relation to the results of Winter \emph{et~al}.~\cite{Winter2010}
we find that the correlation length exponent has been measured in
the regime $29\lesssim z\lesssim239$, leading to the anisotropy exponent
$\theta\approx3$ in agreement with our results. In Ref.~\cite{Gonnella2009}
the correlation length exponents have also been determined in the
mean field limit. Looking at the lowest velocity $v=1/32$ we find
$53\lesssim z\lesssim1066$, where a surprisingly small anisotropy
exponent $\theta\approx0.73$ has been estimated. The highest velocity
$v=50$ leads to $\theta\approx1.2$ and $1100\lesssim z\lesssim22000$.
These discrepancies might be attributed to the fact that an integral
quantity, the order parameter, has been measured, as well as to strong
surface effects induced by the open boundary conditions used in the
$\perp$ direction.

\section{Conclusion}

In this work we investigated two recently proposed driven Ising models
with friction due to magnetic interactions, namely the 1+1d and 2d
model, using MC simulations as well as analytical methods. At first
we focused on the strongly anisotropic critical behavior and calculated
the anisotropy exponent $\theta$ in the limit of high driving velocity
$v\rightarrow\infty$. Therefore the perpendicular correlation function
of a cylinder-like geometry was calculated at criticality for different
system sizes. Evaluating the connection between system size and correlation
length, Eq.~(\ref{eq:Lp_vs_xs}), we were able to find the critical
exponents $\theta=3$ as well as $\nu_{\parallel}=3/2$ and $\nu_{\perp}=1/2$.
The analytic deviation of these exponents within the framework of
a Ginzburg-Landau-Wilson Hamiltonian led to the same values. Comparing
the results to the driven lattice gas \cite{Katz1983,SchmittmannZia1995}
we note that it also shows a strongly anisotropic phase transition
at a critical temperature which grows with the velocity. Remarkably
this phase transition is characterized by the same critical exponents
at large fields.

Finally we focused on the critical behavior for finite velocities
$v$ and performed extensive MC simulations in order to calculate
the crossover scaling function describing the crossover from the Ising
universality class at $v=0$ to the non-equilibrium critical behavior
at $v\to\infty$. The analysis has exemplarily been done for the 1+1d
model, but as shown, both models belong to the same universality class
and similar results are expected for the 2d model. In the analysis
an additional complexity arised due to the strongly anisotropic characteristics
of the correlations. Therefore we calculated the correlation length
in a cylindrical system, circumventing intricate shape effects. We
were able to identify a crossover length $\Lc(v)$ using a simple
method based on the rescaling of data for each velocity such that
a data collapse occurs. This procedure leads to an excellent data
collapse of all simulation results for different velocities $v$ and
system sizes $\Lp$. 

It turns out that for all finite velocities $v>0$ the models undergo
a crossover, at crossover length $\Lc(v)$, from an quasi-equilibrium
isotropic Ising-like phase transition to a non-equilibrium mean-field
behavior with strongly anisotropic correlations.
\begin{acknowledgments}
We thank Felix M. Schmidt and Matthias Burgsmüller for valuable discussions.
This work was supported by CAPES--DAAD through the PROBRAL program
as well as by the German Research Society (DFG) through SFB 616 ``Energy
Dissipation at Surfaces''. 
\end{acknowledgments}
\appendix

\section{Scaling exponents of the GLW model \label{sec:appendix}}

The following calculation is similar to \cite{BrezinZinnJustin1985}.
Discretizing the integral
\begin{equation}
\beta\bar{\mathcal{H}}=\Lp\int_{0}^{\Ls}\!\!\!\!\mathrm{d}\xs\left(\frac{t}{2}m(\xs)^{2}+\frac{1}{2}m'(\xs)^{2}+\frac{u}{4!}m(\xs)^{4}\right)\label{eq:Hamiltonian_GLW-1}
\end{equation}
 with step size $\dx$, $N\dx=\Ls$, $m_{i}=m(i\,\dx)$ and $\delta m_{i}=m_{i+1}-m_{i}$
gives
\begin{equation}
\beta\bar{\mathcal{H}}=\Lp\sum_{i=1}^{N}\dx\left(\frac{t}{2}m_{i}^{2}+\frac{1}{2}\frac{\delta m_{i}^{2}}{\dx^{2}}+\frac{u}{4!}m_{i}^{4}\right).\label{eq:Hamiltonian_GLW-1-1-1}
\end{equation}
In order to evaluate the partition function 
\begin{equation}
\mathcal{Z}=\C{\int_{-\infty}^{\infty}\left(\prod_{\xs=0}^{\Ls}\mathrm{d}m(\xs)\right)e^{-\beta\bar{\mathcal{H}}}=}\int_{-\infty}^{\infty}\mathcal{D}[m(\xs)]\, e^{-\beta\bar{\mathcal{H}}},\label{eq:partition-function-1}
\end{equation}
we use abbreviations in analogy to transfer matrices,
\begin{equation}
T(m,m^{+})=\underbrace{\vphantom{{\textstyle \sqrt{\frac{\Lp}{2\pi\dx}}}}e^{-\Lp\dx\left({\textstyle \frac{t}{2}}m^{2}+{\textstyle \frac{u}{4!}}m^{4}\right)}}_{V(m)}\,\underbrace{{\textstyle \sqrt{\frac{\Lp}{2\pi\dx}}}e^{-{\textstyle \frac{\Lp\delta m^{2}}{2\dx}}}}_{U(\delta m)},
\end{equation}
with $m^{+}=m+\delta m$ to get 
\begin{eqnarray}
\mathcal{Z} & = & \int_{-\infty}^{\infty}\negthickspace\negthickspace\negthickspace\mathrm{d}m_{1}\int_{-\infty}^{\infty}\negthickspace\negthickspace\negthickspace\mathrm{d}m_{2}T(m_{1},m_{2})\int_{-\infty}^{\infty}\negthickspace\negthickspace\negthickspace\mathrm{d}m_{3}T(m_{2},m_{3})\times\nonumber \\
 &  & \cdots\times\int_{-\infty}^{\infty}\negthickspace\negthickspace\negthickspace\mathrm{d}m_{N}T(m_{N-1},m_{N})T(m_{N},m_{1})
\end{eqnarray}
for the assumed periodic boundary conditions. 

Let $\xs^{+}=\xs+\dx$ and $\f(m^{+})$ be the result of the integrations
for the interval $]\xs^{+},\Ls]$. Since $T(m,m^{+})$ is near-diagonal
for $\Lp\rightarrow\infty$, we can write $\f(m^{+})$ as 
\begin{equation}
\lambda\f(m^{+})\approx\f(m)+\f'(m)\delta m+\frac{1}{2}\f''(m)\delta m^{2},
\end{equation}
where $\lambda$ denotes the growth factor of the integrations corresponding
to the leading eigenvalue of the transfer matrix $T(m,m^{+})$. The
integral over $m^{+}$ in the partition function becomes 
\begin{eqnarray}
\f(m) & = & \int_{-\infty}^{\infty}\negthickspace\negthickspace\negthickspace\mathrm{d}m^{+}V(m)U(m^{+}-m)\f(m^{+})\nonumber \\
 & = & V(m)\sqrt{\frac{\Lp}{2\pi\dx}}\int_{-\infty}^{\infty}\negthickspace\negthickspace\negthickspace\mathrm{d}m^{+}e^{-{\textstyle \frac{\Lp\delta m^{2}}{2\dx}}}\f(m^{+})\nonumber \\
 & = & \frac{V(m)}{\lambda}\left(\f(m)+\frac{\dx}{2\Lp}\f''(m)\right),
\end{eqnarray}
and yields the solution of the integrations for the interval $[\xs^{+},L_{\perp}]$.
Hence we get a differential equation for $\f(m)$, 
\begin{equation}
V(m)\left(\f(m)+\frac{\dx}{2\Lp}\f''(m)\right)=\lambda\f(m).\label{eq:dgl}
\end{equation}
We now substitute \bS\label{eq:Substitution} 
\begin{eqnarray}
\f(m) & \:\rightarrow\: & \Psi(\tilde{m})\\
m & \:\rightarrow\: & \tilde{m}\, u^{-1/6}\Lp^{-1/3}\label{eq:rescal_m}\\
\lambda & \:\rightarrow\: & 1-\Lambda\,\dx\, u^{1/3}\Lp^{-1/3}\\
t & \:\rightarrow\: & x\, u^{2/3}\Lp^{-2/3}
\end{eqnarray}
\eS and expand to lowest order around $\Lp=\infty$ to yield the
Schrödinger equation in a quartic potential, 
\begin{equation}
\left(-\frac{1}{2}\partial_{\tilde{m}}^{2}+\frac{x}{2}\tilde{m}^{2}+\frac{1}{4!}\tilde{m}^{4}-\Lambda\right)\Psi(\tilde{m})=0,\label{eq:schrodinger}
\end{equation}
valid in the scaling limit $\Lp\to\infty$, $t\to0$ with $x=t(\Lp/u)^{1/\nu_{\parallel}}$
kept fixed.

The correlation length $\xi_{\perp}(\Lp)$ is determined from the
lowest eigenvalues $\Lambda_{0,1}$ of this equation, as
\begin{equation}
\xi_{\perp}=\dx\left(\log\frac{\lambda_{0}}{\lambda_{1}}\right)^{-1}\sim\frac{1}{\Lambda_{1}-\Lambda_{0}}\left(\frac{\Lp}{u}\right)^{1/3}.\label{eq:}
\end{equation}
From the substitution, Eqs.~(\ref{eq:Substitution}), we directly
read off the exponents $\nu_{\parallel}=3/2$, and $\theta=3$. 

The correlation function amplitude $\hat{G}_{\perp}(\Lp)$ from Eq.~(\ref{eq:Gperp})
is proportional to $m^{2}$ and thus scales as $\Lp^{-2/3}$ as can
be seen from Eq.~(\ref{eq:rescal_m}).

\bibliographystyle{apsrev4-1}
\bibliography{Literatur}

\end{document}